\pdfoutput=1
\documentclass[12pt]{article}
\usepackage{authblk}
\usepackage{amssymb,amsmath,bm}
\usepackage{mathtools}
\mathtoolsset{showonlyrefs}

\usepackage{amsthm,,amsfonts,caption}

\usepackage[bottom]{footmisc}

\usepackage{newtxtext}       %
\usepackage[varvw]{newtxmath}       

\usepackage{appendix}
\usepackage{graphicx}
\newcommand{\bX}{\mathbf X}
\newcommand{\bfx}{\mathbf{x}}
\newcommand{\bfX}{\mathbf{X}}
\newcommand{\Ent}{\text{Ent}}
\newcommand{\Argmax}{\mathop{\mathrm{argmax}}}

\def\N{\mathbb N}
\def\R{\mathbb R}
\def\eps{\varepsilon}

\def\H {\operatorname{H}}

\newcommand{ \ep }{\varepsilon}
\newcommand{\ent}{{\rm Ent}}        
\newcommand{\suchthat}{\ensuremath{\ : \ }} 
\newcommand{\Prob}{\mathcal P}

\newcommand{\Tr}{\operatorname{Tr}}
\newcommand{\DM}{\mathfrak{P}}
\newcommand{\unit}{1\!\!1}
\newcommand{\prim}{\mathfrak{F}^{\ep}}

\newcommand{\primin}{\mathfrak{F}^{\ep}}
\newcommand{\dual}{\mathfrak{D}^{\ep}}
\newcommand{\dualf}{\tD_{\bm \gamma}^\ep}
\newcommand{\dualin}{\mathfrak{D}^{\ep}}

\newcommand{\Sink}{\mathcal{T}^\ep}
\newcommand{\ffD}{\text D_\gamma^{-,\ep}}
\newcommand{\ffd}{\text D_\gamma^{-,\eps}}
\newcommand{\fbD}{\text D_\gamma^{+, \ep}}

\newcommand{\fbfD}{\text D_\gamma^{\pm, \ep}}

\newcommand{\fD}{\mathfrak D^\ep_-}
\newcommand{\bD}{\mathfrak D^\ep_+}
\newcommand{\bfD}{\mathfrak D_\pm^\ep} 

\newcommand{\tand}{\quad \text{and} \quad}

\theoremstyle{plain}
\newtheorem{theo}{Theorem}[section]
\newtheorem{proposition}[theo]{Proposition}

\theoremstyle{definition}

\newtheorem{rem}[theo]{Remark}

\newcommand{\cM}{\mathcal{M}}

\newcommand{\cS}{\mathcal{S}}

\newcommand{\cK}{\mathcal{K}}

\newcommand{\bR}{\mathbb{R}}

\newcommand{\bC}{\mathbb{C}}

\newcommand{\tP}{\operatorname{P}}

\newcommand{\tD}{\operatorname{D}}

\newcommand{\tS}{\operatorname{S}}
\newcommand{\ren}{\operatorname{Ren}}

\newcommand{\fT}{\mathfrak{T}}
\newcommand{\fg}{\mathfrak g}
\newcommand{\fh}{\mathfrak h}

\DeclareFontFamily{U}{mathx}{\hyphenchar\font45}
\DeclareFontShape{U}{mathx}{m}{n}{
	<5> <6> <7> <8> <9> <10>
	<10.95> <12> <14.4> <17.28> <20.74> <24.88>
	mathx10
}{}
\DeclareSymbolFont{mathx}{U}{mathx}{m}{n}

\newcommand{\dd}{\, \mathrm{d}}

\title{Entropic regularised optimal transport in a noncommutative setting}

\author{Lorenzo Portinale}
\affil{Institut f\"ur Angewandte Mathematik, Universit\"at Bonn, Germany}
\date{}

\begin{document}
\maketitle
\begin{abstract}{
	This survey has been written in occasion of the School and Workshop about Optimal Transport on Quantum Structures at Erdös Center in September 2022.
	We discuss some recent results on noncommutative entropic optimal transport problems and their relation to the study of the ground-state energy of a finite-dimensional composite quantum system at positive temperature, following the work \cite{Feliciangeli-Gerolin-Portinale:2022}.
	In the first part, we review some of the classical primal-dual formulations of optimal transport in the commutative setting, including extensions to multimarginal problems and entropic regularisation. We discuss the main features of the entropic problem and show how optimisers can be efficiently computed via the so-called Sinkhorn algorithm.
	In the second part, we discuss how to apply these ideas to a noncommutative setting, in particular on the space of density matrices over finite dimensional Hilbert spaces. In this framework, we present equivalences between primal and dual formulations, and use them to characterise the optimisers.
	Despite the lack of explicit formulas due to the noncommutative nature of the problem, one can also show that a suitable quantum version of the Sinkhorn algorithm converges to the minimiser of the entropic problem. 
	In the final part of this work, we discuss similar results for bosonic and fermionic systems.}
\end{abstract}
\section{Introduction}
The goal of this paper is to discuss how the ideas from the theory of classical optimal transport can be used and transferred to the study of variational problems in noncommutative analysis. In particular, we focus on the study of entropy-regularised transport problems (for a classical survey on the topic, see \cite{LeoSurvey}) and their dual formulations, whose particular structure provides the existence of a fast algorithm to compute the corresponding optimisers. This is called \textit{Sinhorn's algorithm} and its origin goes way back to \cite{Sin64}. In the context of Computational Optimal Transport, the entropic regularization and the Sinkhorn algorithm were introduced and studied in \cite{Cut13, GalSal}.

In this work, we are going to review the main features of the Sinkhorn algorithm in the classical setting, and discuss how to extend its use and application to the setting of quantum optimal transport, following the recent work \cite{Feliciangeli-Gerolin-Portinale:2022}. Using a more physical point of view, a noncommutative version of the entropy-regularised optimal transport problem naturally arises within the study of the ground-state energy of a finite-dimensional composite quantum system at positive temperature. In other words, the aim is to look for the minimal energy levels of a composite system conditionally to the knowledge of the states of all its subsystems. Or, in similar terms: which information can we gather on the state of a composite system when we only have access (e.g. experimentally) to measurements of the states of its subsystems?

In the final part of the paper, we also discuss the problem when enforcing symmetry conditions, or more precisely, the \textit{fermionic} and \textit{bosonic} framework. This creates a bridge to what in the physics literature is known as One-body Reduced Density Matrix Functional Theory ($1$RDMFT), introduced in 1975 by Gilbert \cite{Gil75} as an extension of the Hohenberg-Kohn (Levy-Lieb) formulation of Density Functional Theory (DFT) \cite{HohKoh,Levy,Lieb83}, and which can be framed as a special case of \cite{Feliciangeli-Gerolin-Portinale:2022}. 

\smallskip
\noindent 
\textit{Some literature}: \
The work \cite{Feliciangeli-Gerolin-Portinale:2022} is not the first attempt to apply the ideas of optimal transport to the noncommutative setting. One of the first attempt was carried out by E. Carlen and J. Maas \cite{CarMas14}, followed by many others. It is important to observe a fundamental difference between two approaches: commutative optimal transport can be cast \textit{equivalently} as a static coupling problem or as a dynamical optimisation problem. In the noncommutative setting, the relation (if any) between the two interpretations is not clear. This draws a line between the two groups of work, the ones that consider the dynamical formulations (e.g. \cite{CarMas14, CheGeoan18, MitMie17, MonVor20}) and the ones which instead focus on a static formulation (e.g. \cite{CalGolPau18, DPaTre19, GeoPav15, MitMie17, PeyChiViaSol19}). 
Concerning duality formulations in the quantum setting, in \cite{CalGolPau18} the authors study the case of $\eps=0$ and prove a duality result for the noncommutative problem in the very same spirit of the Kantorovich duality for the classical Monge problem. In the dynamical framework, \cite{Wirth:2021} studies the entropic quantum optimal transport problem as well, in constrast to the static approach adopted in \cite{Feliciangeli-Gerolin-Portinale:2022}. 
For more details on the state of the art, both of quantum optimal transport and 1RDMFT, see \cite{Feliciangeli-Gerolin-Portinale:2022} and references therein.

\smallskip
\noindent 
\textit{Organisation of the paper}: \
In the first section we review part of the classical theory of (regularised) optimal transport problems. We then move to the noncommutative framework, focusing on similarities and differences with the classical setting. Subsequently, we introduce the noncommutative Sinkhorn algorithm and discuss the convergence toward optimisers (both in the primal and dual problem). At last, we present the fermionic/bosonic setting, and discuss a variational interpretation of the Pauli exclusion principle.

\section{Classical Optimal Transport and Regularisation}
The classical optimal transport problem consists in minimising the total cost of transportation of a given probability measure $\mu \in \Prob(X)$ into a target one $\nu \in \Prob(Y)$, with respect to some cost function $c:X \times Y \to \R$, which for simplicity here we assume to be continuous and bounded, i.e. $c \in C_b(X\times Y)$. In the most general setting, $X$, $Y$ are Polish spaces, i.e. topological spaces that can be endowed with a distance which makes them complete and separable. We refer to the seminal book \cite{VilON} for a general treatment of this subject. The \textit{Kantorovich} formulation of the associated optimal transport problem then reads as
\begin{align}
	\label{eq:classic_OT}
	\inf_{\pi \in \Prob(X \times Y)}
	\left\{
		\int c(x,y) \dd \pi(x,y) 
			\suchthat 
		\big( \text{Proj}_X \big)_{\#} \pi = \mu 
			\tand 
		\big( \text{Proj}_Y \big)_{\#} \pi = \nu 
	\right\} \, , 
\end{align}
where $\text{Proj}_X: (X,Y) \to X$ (resp. $\text{Proj}_Y: (X,Y) \to Y$) denotes the projection operator onto $X$ (resp. Y). A probability measure $\pi \in \Prob(X\times Y)$ which satisfies the marginal constraint as in \eqref{eq:classic_OT} is called an \textit{admissible transport plan} between $\mu$ and $\nu$. When the involved marginal measures are atomic (or discrete), i.e. of the form
\begin{align}
	\mu = \sum_{i=1}^L \mu_i \delta_{x_i}
		\tand 
	\nu = \sum_{j=1}^M \nu_j \delta_{y_j} \, , 
\end{align}
then the variational problem \eqref{eq:classic_OT} is a fully discrete constrained linear program which can be equivalently recast as
\begin{align}
	\label{eq:discrete_classic_OT}
	\inf_{\pi_{ij} \geq 0}
	\left\{
		\sum_{i=1}^L \sum_{j=1}^M
			c(x_i,y_j) \pi_{ij}
	\suchthat 
		\sum_{j=1}^M \pi_{ij} = \mu_i 
			\tand 
		\sum_{i=1}^L \pi_{ij} = \nu_j 
	\right\} \, .
\end{align}
An important feature of the variational problem \eqref{eq:classic_OT} is the fact it admits a \textit{dual formulation}, using the natural duality between the space of measures and the space continuous and bounded functions. 
The equivalent, dual formulation of \eqref{eq:classic_OT} is given by
\begin{align}	\label{eq:dual_classic}
	\sup_{\varphi, \psi} 
	\left\{
		\int \varphi(x) \dd \mu(x)
			+ 
		\int \psi(y) \dd \nu(y)	
			\suchthat 
		\varphi(x) + \psi(y) \leq c(x,y)
	\right\} \, ,
\end{align}
where the supremum is taken over $\varphi \in C_b(X)$ and $\psi \in C_b(Y)$.

As we are going to see several times in this survey, a natural technique in order to look for maximising couples $(\varphi,\psi)$ in \eqref{eq:dual_classic} consists in fixing one of two functions and maximising the target functional with respect to the second one. In more precise terms, for a given $\varphi \in C_b(X)$, we define $\varphi^c \in C_b(Y)$ as the solution to
\begin{align}
	\varphi^c(y) := \inf_{x \in X} \big\{ c(x,y) - \varphi(x) \big\} = \Argmax_\psi
	\left\{
		\int \psi(y) \dd \nu(y) 
			\suchthat 
		\varphi(x) + \psi(y) \leq c(x,y)
	\right\} \, .
\end{align}
We call $\varphi^c$ the \textit{$c$-transform} of $\varphi$. Similarly, for any given $\psi \in C_b(Y)$, one defines $\psi^c \in C_b(X)$ by maximising \eqref{eq:dual_classic} over $\varphi$, or equivalently
\begin{align}
	\psi^c(x) := \inf_{y \in Y} 
	\big\{
		c(x,y) - \psi(y)
	\big\}	 \in C_b(X) \, .
\end{align} 
We say that $\psi$ is \textit{c-concave} if $\psi= \varphi^c$ for some $\varphi$. One can prove that $\varphi^{ccc} = \varphi^c$ and that $\psi$ is c-concave if and only if $\psi^{cc} = \psi$ (see e.g. \cite[Prop.~5.8]{VilON}). This means that the optimisers in \eqref{eq:dual_classic} are necessarily a c-concave function and its c-trasform, or in other words we can write \eqref{eq:dual_classic} equivalently as
\begin{align}
	\sup_{\varphi \in C_b(X)} 
		\left\{
			\int \varphi(x) \dd \mu(x)
				+ 
			\int \varphi^c(y) \dd \nu(y)	
				\suchthat 
			\varphi \text{ is c-concave}
		\right\} \, .
\end{align}
The optimal $\varphi$ in the dual problem is usually called \textit{Kantorovich potential}.

\begin{rem}[Euclidean spaces] 
	\label{rem:euclidean_concentration}
A widely studied example of optimal transport is the one corresponding to the euclidean space $X = Y =\R^d$ and the quadratic cost $c(x,y) = |x-y|^2$. In this case, the square root of \eqref{eq:classic_OT} defines a distance on the space of probability measures with finite second moment known in literature as \textit{Kantorovich--Rubenstein--Wasserstein} distance $W_2$. One of the consequence of the duality formula \eqref{eq:dual_classic} (and in particular the characterisation of $c$-concave maps) in this framework is the existence of the so-called \textit{optimal transport map} $T: \R^d \to \R^d$, whenever $\mu$ is absolutely continuous with respect to the Lebesgue measure in $\R^d$. This means that the optimal plan $\pi$ in \eqref{eq:classic_OT} is of the form $\pi = (\text{id}, T)_{\#} \mu$, which corresponds to the original formulation of the problem, as introduced by Monge \cite{Monge} back in the 18th century. In particular, $\pi$ is supported on the graph of the map $T$. Note that this phenomenon is quite in contrast with what happens in the fully discrete case \eqref{eq:discrete_classic_OT}, where in most cases no admissible plan is induced by a map $T$.
\end{rem}

\subsubsection*{Multimarginal Optimal transport}
Up to this point, we have discussed two marginals problems, namely where the optimal transport problem happens between two spaces and two probability measures $\mu$ and $\nu$. A natural, interesting generalisation are the \textit{multimarginal optimal transport problems}, where instead of an initial and a target space, one considers $N$ spaces $X_1, \dots, X_N$ and $N$ given marginals $\{\mu_i \in \Prob(X_i)\, : \, i= 1, \dots, N\}$. This time a cost function is given by a map from the product of all the spaces, namely
\begin{align}
	\bX:=\bigtimes_{i=1}^N X_i
		\tand 
	c: \bX \to \R 
	\quad \text{continuous and bounded} \, .
\end{align}
Using the notation $\bfx:= (x_1, \dots, x_N)$, the variational problem associated to $c$ and $\{\mu_i\}_i$ is then given by
\begin{align}
\label{eq:multim_classic_OT}
	\inf_{\pi \in \Prob(\bX)}
	\left\{
		\int c(\bfx) \dd \pi(\bfx) 
			\suchthat 
		\big( \text{Proj}_{X_i} \big)_{\#} \pi = \mu_i 
			\quad \text{for} \ i = 1, \dots, N 	
	\right\} \, .
\end{align}

In this picture, we lose the original interpretation of transport from a given origin to a target location. Nonetheless, in some special cases, one can reduce this formulation to the "Monge" one, by considering admissible multimarginal plans associated with $N-1$ maps, of the form $\pi = (\text{id}, T_1, \dots, T_{N-1})_{\#} \mu_1$, see e.g. \cite{Gangbo-Swicech:1998}. In a similar spirit as for the two-marginals case, the primal problem \eqref{eq:multim_classic_OT} admits a dual formulation, which is given by
\begin{align}
	\label{eq:multim_dual_classic}
	\sup
	\left\{
		\sum_{i=1}^N
			\int_{X_i} \varphi_i(x_i) \dd \mu_i(x_i)
	\suchthat 
		\varphi_j  \in C_b(X_j)
			\tand
		\sum_{i=1}^N \varphi_i(x_i) \leq c(\bfx)
	\right\} \, .
\end{align}
Moreover, we can define the associated $c$-transform operators: this time, they take as input $N-1$ functions, and find the optimal $N$-th function that satisfies the cost constraint. More precisely, for a fixed index $j \in \{1, \dots, n\}$, the $c$-transform of the continuous functions $(\varphi_1, \dots, \varphi_{j-1}, \varphi_{j+1}, \dots, \varphi_N)$ is the continuous function given by
\begin{align}	\label{eq:def_ctrasnf_multimarginal}
	(\varphi_1, \dots, \varphi_{j-1}, \varphi_{j+1}, \dots, \varphi_N)^c(x_j):= \inf_{\mathbf{x}^{(j)}}
	\left\{
		c(\bfx) - \sum_{i \neq j} \varphi_i(x_i)
	\right\} \, ,
\end{align}
where we used the notation $\mathbf{x}^{(j)}:= (x_1, \dots, x_{j-1}, x_{j+1}, \dots, x_N) \in \bigtimes_{i \neq j} X_i$. 

\subsubsection*{Regularisation of optimal transport}
Especially in the two marginal case, the optimal transport plans can be very singular with respect to diffused measures on the product, for example the Lebesgue measure on $\R^d \times \R^d$. They are instead often concentrated on the graph of functions, and therefore less "diffused" (cfr. Remark~\ref{rem:euclidean_concentration}). In order to obtain less singular optimisers, one possibility is to introduce a penalisation term in the minimisation that forces the plan to be absolutely continuous with respect to a given reference measure. A typical energy that penalises singular measures with respect to a given reference is the \textit{relative entropy} functional. For $\sigma \in \Prob(\bfX)$ a given reference measure, the relative entropy w.r.t. $\sigma$ is defined as
\begin{align}
	\label{eq:def_entropy}
	\Ent(\cdot|\sigma): \Prob(\bfX) \to [0,+\infty]
		\, , \qquad 
	\Ent(\pi|\sigma) := 
	\begin{cases}
		\int \rho \log \rho \dd \sigma 
			&\text{if } \dd \pi= \rho \dd \sigma \, , 
	\\
		+ \infty
			&\text{otherwise} \, . 
	\end{cases}
\end{align}
In particular, if $\ent(\pi|\sigma) < \infty$, then $\pi$ is necessarily absolutely continuous with respect to $\sigma$. Note that we are adopting the mathematical convention, which means that the entropy of a probability measure is a positive number (and not negative, as more commonly assumed in physics).
 For a given parameter $\eps \in (0,+\infty)$, we can then define a $\eps$-regularised multimarginal optimal transport problem by adding an entropy penalisation, that is
\begin{align}
\label{eq:multim_regularised_OT}
	\inf_{\pi \in \Prob(\bX)}
	\left\{
		\int c(\bfx) \dd \pi(\bfx) 
			+ \eps \Ent(\pi|\sigma)
			\suchthat 
		\big( \text{Proj}_{X_i} \big)_{\#} \pi = \mu_i 
			\quad \text{for} \ i = 1, \dots, N 	
	\right\} \, .	
\end{align}
In order to have a well-posed problem, we assume $\sigma = \bigotimes_i \mathfrak{m}_i$ and $\Ent(\mu_j| \mathfrak{m}_j)<\infty$, for every $j$.
Note that the minimiser of the regularised problem \eqref{eq:multim_regularised_OT} with $\eps>0$ must have finite entropy, therefore it is necessarily absolutely continuous with respect to $\sigma$. Furthermore, by means of a simple computation, we see that 
\begin{align}
	\int c(\bfx) \dd \pi(\bfx) 
		+ \eps \Ent(\pi|\sigma)
&=
	\eps \int \rho \big( \log e^{\frac{c}\eps} + \log \rho \big) \dd \sigma
\\
&=
	\eps \int e^{\frac{c}\eps}\rho  \log (\rho e^{\frac{c}\eps} ) e^{-\frac{c}\eps} \dd \sigma
= \eps\Ent(\pi|e^{-\frac{c}\eps} \sigma) \, ,
\end{align}
which in  particular shows that we can equivalently recast \eqref{eq:multim_regularised_OT} as 
\begin{align}
	\label{eq:multim_Schrodinger}
		\inf_{\pi \in \Prob(\bX)}
		\left\{
			\Ent(\pi|e^{-\frac{c}\eps} \sigma)
				\suchthat 
			\big( \text{Proj}_{X_i} \big)_{\#} \pi = \mu_i 
				\quad \text{for} \ i = 1, \dots, N 	
		\right\} \, .
\end{align}
In other words, the $\eps$-regularised optimal transport problem is equivalent to finding the minimiser of the relative entropy with respect to the weighted measure $\mathcal K_\eps:= e^{-\frac{c}\eps}\sigma$. This is known in the mathematical literature as \textit{Schr\"odinger problem} associated with $c$ and $\sigma$, and $\mathcal K_\eps$ is often referred to as \textit{Gibbs measure}.

Even with a positive regularisation parameter $\eps>0$, the problem \eqref{eq:multim_regularised_OT} admits a dual formulation. This time, the presence of a superlinear term (the entropy) causes a particular phenomenon: the dual problem is not a linearly constrained problem (as for $\eps=0$) where the cost appears in the constraint. On the contrary, it is obtained as a \textit{free maximisation problem} where the marginals appear linearly in the dual functional, whereas the cost appears as part of a (concave) nonlinear term. The dual formulation of \eqref{eq:multim_regularised_OT} indeed reads as
\begin{align}
	\label{eq:multim_regularised_dual_OT}
	\sup
	\left\{
		\sum_{i=1}^N
			\int_{X_i} \varphi_i(x_i) \dd \mu_i(x_i)
	- \eps 
		\int_{\bfX}
		\exp
		\left\{
			\frac1\eps 
			\bigg(
				\sum_{i=1}^N \varphi_i(x_i) - c(\bfx)		
			\bigg)
		\right\}
				\dd \sigma(\bfx)
	\right\} + \eps \, , \
\end{align}
where the supremum runs over all $\varphi_i  \in C_b(X_i)$. At least formally, by sending $\eps \to 0$, we recover the dual formula \eqref{eq:multim_dual_classic}. This is related to the so-called \textit{Laplace principle}, which informally states that
\begin{align}
\int_{\bfX}
	\exp
		\left\{
			\frac1\eps 
			\bigg(
				\sum_{i=1}^N \varphi_i(x_i) - c(\bfx)		
			\bigg)
		\right\}
				\dd \sigma(\bfx)
	\underset{\eps \to 0}{\approx}
	\exp
		\left\{
			\frac1\eps
			\sigma
				\text{-esssup} 
			\bigg(
				\sum_{i=1}^N \varphi_i(x_i) - c(\bfx)		
			\bigg)
		\right\} \, ,
\end{align}
showing that, in the limit as $\eps \to 0$, in order to have a competitor which has energy value which is not $-\infty$, one must necessarily enforce the constraint as in \eqref{eq:multim_dual_classic}.
\begin{rem}[Schr\"odinger problem in $\R^d$]
Let us consider the Euclidean and two-marginals setup, i.e. $N=2$, $X = Y = \R^d$, and consider $\sigma:= \dd x \dd y$ the Lebesgue measure on the product, together with the quadratic cost (up to translation by a constant) $c(x,y)= \frac12|x-y|^2 - \log \sqrt{2 \pi \eps^2}$. In this case, the Gibbs measure coincides with the Gaussian kernel $\mathcal K_\eps = \frac1{\sqrt{2 \pi \eps^2}} e^{-\frac{|x-y|^2}{2\eps}}\dd x \dd y$ and
	\begin{align*}
		\inf_{\pi}
		\big\{
			\Ent \big( \pi | \mathcal K_\eps	\big) 
			\suchthat 
			\big( \text{Proj}_{X_i} \big)_{\#} \pi = \mu_i 
		\big\}
			= 
		\inf_P
		\big\{
		 \Ent 
			\big(
				P | W_\eps
			\big)
			\suchthat P(\cdot | t=i) = \mu_i 
		\big\}\, ,
	\end{align*}
	where $P \in \Prob \big( C([0,1]; \R^d) \big)$ and $W_\eps\in\Prob \big( C([0,1]; \R^d) \big)$ is the Wiener measure.
This represents a bridge between the theory of (regularised) optimal transport and \textit{Large Deviations}. Indeed, as described by Sanov's Theorem, the relative entropy functional $\Ent(\cdot | W_\eps)$ naturally appears as rate functional for the large deviation principle satisfied by i.i.d. random variables equally distributed as $W_\eps$. 
\end{rem}

As a typical feature of primal-dual problems, we have that the minimisers and maximisers of the respective variational problems are related to each other. In this case, we can write an explicit formula for the minimal coupling $\pi_\eps \in \Prob(\bfX)$ in \eqref{eq:multim_regularised_OT} in terms of the optimal Kantorovich potentials $(\varphi_{\eps,1}, \dots, \varphi_{\eps,N})$, $\varphi_{\eps,i} \in C_b(X_i)$ in \eqref{eq:multim_regularised_dual_OT}. It turns out that $\pi_\eps$ is given the absolutely continuous measure with respect to $\cK_\eps$ with density given by
\begin{align}
	\label{eq:optimiser_formula}
	\frac
		{\dd \pi_\eps}
		{\dd \cK_\eps}
	(\bfx) 
		= 
	\exp
	\left(
		\frac1\eps 
		\left(
			\sum_{i=1}^N \varphi_{\eps,i}(x_i) - c(\bfx)
		\right) 
	\right) 
=
	\prod_{i=1}^{N}
		\exp
		\left(
			\frac
				{\varphi_{\eps,i}(x_i) - N^{-1}c(\bfx)}{\eps}
		\right)
		\, .
\end{align}
In particular, once evaluated on $\pi_\eps$, the nonlinear (exponential) integral term in \eqref{eq:multim_regularised_dual_OT} simply gives $1$ (cause $\pi_\eps$ is a probability measure) and value of the primal \eqref{eq:multim_regularised_OT} and dual \eqref{eq:multim_regularised_dual_OT} problem coincides, and it is given by
\begin{align}
	\sum_{i=1}^N \int_{X_i} \varphi_{\eps,i}(x_i) \dd \mu_i(x_i) 
		=
	\int c(\bfx) \dd \pi_\eps(\bfx) 
	+ \eps \Ent(\pi|\sigma) \, .
\end{align}

The entropic regularised problem also admits a natural notion of $c$-transform, that the authors in \cite{DMaGer19} call \textit{$(c,\eps)$-trasform}. In a similar way as for the case $\eps=0$, it is a trasformation which corresponds to optimising over a \textit{single} potential, say $\varphi_j$, while fixing the other $N-1$, that is $\{\varphi_i \, : \, i \neq j\}$. In the regularised setting ($\eps>0$), one can write explicitly the solution to this maximisation problem. In particular, for a fixed index $j \in \{1, \dots, n\}$, the $(c,\eps)$-transform of the continuous functions $(\varphi_1, \dots, \varphi_{j-1}, \varphi_{j+1}, \dots, \varphi_N)$ is the continuous function denoted by $	(\varphi_1, \dots, \varphi_{j-1}, \varphi_{j+1}, \dots, \varphi_N)^{(c,\eps)}$ and given by
\begin{align}	\label{eq:def_ctrasnf_multimarginal_regularised}
	x_j \mapsto   
-\ep\log\left(\int_{\bigtimes_{i \neq j}X_i} \exp\left(\frac1{\ep}
\left( 
\sum_{i\neq j}\varphi_i(x_i)-c(\bfx)
\right)
\right){\rm d} \bigotimes^N_{i\neq j}\mathfrak{m}_i  \right) + \ep\log \Big( \frac{\dd \mu_j}{\dd \mathfrak{m}_j} \Big) 
		\, .
\end{align}
Once again, by the Laplace principle, in the limit as $\eps \to 0$, one recovers (at least formally) the definition of $c$-transform as given in \eqref{eq:def_ctrasnf_multimarginal}.

\subsubsection*{Sinkhorn algorithm}
In order to describe the Sinkhorn algorithm, let us start the discussion in the case of the two-marginals problems. Namely, for $\eps>0$, we seek solutions to 
\begin{align}
	\inf_{\pi \in \Prob(X \times Y)}
	\left\{
		\Ent \big( \pi | \mathcal K_\eps	\big) 
		\suchthat 
		\big( \text{Proj}_X \big)_{\#} \pi = \mu
			\tand
		\big( \text{Proj}_Y \big)_{\#} \pi = \nu 
	\right\}
		\, , \quad 
		\mathcal K_\eps := e^{\frac{-c}{\eps}} \dd \sigma \, .
\end{align}
The key idea behind Sinkhorn algorithm is to solve alternating optimisations problems in order to find a sequence of plans $\{ \pi^{(n)} \}_n$ having one correct marginal at a time, in such a way that $\pi^{(n)} \to \pi_\eps$ (the optimal solution) as $n \to \infty$.
Moreover, from the optimality conditions described in \eqref{eq:optimiser_formula}, we know that the optimiser must be in product form. In fact, one can prove that it completely characterises the minimiser of the primal problem: if $\pi_\eps$ has the right marginals, and it is in product form, then it must be optimal. Driven by these considerations, one constructs the following algorithm.

\noindent
{\underline{Initialisation}. \
We initialise $\pi^{(0)}= (f^{(0)} \otimes g^{(0)}) \mathcal K_\eps \in \Prob(X \times Y)$ in product form.  

\noindent
{\underline{Iteration}}. \ 
We iteratively define $\pi^{(n)}= (f^{(n)} \otimes g^{(n)}) \mathcal K_\eps \in \Prob(X \times Y)$ by  imposing the correct marginals $\mu$, $\nu$ one at a time, i.e. given $f^{(n-1)}$, $g^{(n-1)}$, we solve
\begin{align}	\label{eq:iteration_Sinkhorn_twomarginals}
	(\text{Proj}_X)_{\#} (f^{(n)} \otimes g^{(n-1)}) \mathcal K_\eps = \mu 
		\, , \quad 
	(\text{Proj}_Y)_{\#} (f^{(n)} \otimes g^{(n)}) \mathcal K_\eps = \nu
		\, .
\end{align}
Explicitly, for $\dd \mu = \rho \dd \mathfrak m_1$, $\dd \nu = \eta \dd \mathfrak m_2$, the iterations are given by
\begin{align}	\label{explicit}
	f^{(n)}(x) = 
		\frac
			{\rho(x)}
			{ \int_Y g^{(n-1)}(y) e^{-c(x,y)/\eps} \dd \mathfrak m_2(y) }
		\, , \quad 
	g^{(n)}(y) = 
		\frac
			{\eta(y)}
			{ \int_X f^{(n)}(x) e^{-c(x,y)/\eps} \dd \mathfrak m_1(x) }
		\, . \quad 
\end{align}
In this setting, one can prove the following result: assume that $c \in L^\infty(X \times Y)$. Up to subsequence, $f^{(n)}/\lambda^{(n)} \to f_\eps$ in $L^1(\mathfrak m_0)$ and $g^{(n)}/ \lambda^{(n)} \to g_\eps$ in $L^1(\mathfrak m_1)$, for some $\{ \lambda^{(n)} >0\}_n$. In particular, $\pi^{(n)} \to \pi_\eps$ in $L^1(\sigma)$, where $\pi_\eps$ is optimal for \eqref{eq:multim_regularised_OT}. 
The convergence of this procedure was first provided by J. Franklin and J. Lorenz \cite{FraLor89} in the discrete case, via the introduction of a Hilbert metric on the projection cone of the Sinkhorn iterations, and by L. Ruschendorf \cite{ruschendorf1995convergence} in the continuous one. An alternative proof based on the Franklin-Lorenz approach was also provided by Y. Chen, T. Georgiou and M. Pavon \cite{chen2016entropic}, which is particular yields a linear convergence rate of the algorithm (in the Hilbert metric). 

It turns out that solving \eqref{eq:iteration_Sinkhorn_twomarginals} is equivalent to compute the $(c,\eps)$-transform of the previous step. More precisely, one can write 
\begin{align}
	f^{(n)}:= \exp(\varphi^{(n)})
		\, , \ 
	g^{(n)}:= \exp(\psi^{(n)})
		\, , \quad \text{and} \ \ \ 
	\varphi^{(n)} := \big(\psi^{(n-1)}\big)^{(c,\eps)}
		\, , \ 
	\psi^{(n)} := \big(\varphi^{(n)}\big)^{(c,\eps)}
		.
\end{align}
In the work \cite{DMaGer19}, the authors proved the convergence the Sinkhorn algorithm in the setting of multimarginal optimal transport. In this case, one initialise as $\pi^{(0)} = \big( \bigotimes_{j=1}^N f_j \big) \mathcal K_\eps$ and consider the iterations $\pi^{(n)} = \big( \bigotimes_{j=1}^N f_j^{(n)} \big) \mathcal K_\eps$ where the sequence $\{ f_j^{(n)} := \exp(\varphi_j^{(n)}) \, : \, j=1, \dots, N \}_n$ is defined by
\begin{align}
	(\text{Proj}_{X_i})_{\#} 
	\Big(
		\bigotimes_{i=1}^{j}f_j^{(n)} 
		\bigotimes_{i=j+1}^{N}f_j^{(n-1)}
	\Big)
	\mathcal K_\eps = \mu_i
\, \Leftrightarrow \,
	\varphi_j^{(n)} = \big(\varphi_1^{(n)}, ... \, , \varphi_{j-1}^{(n)}, \varphi_{j+1}^{(n-1)}, ... \big)^{(c,\eps)} \, , 
\end{align}
using the notion of $c$-transform in the sense of \eqref{eq:def_ctrasnf_multimarginal_regularised}. In particular, it has been showed in \cite{DMaGer19} that for bounded cost functions, there exist renormalisation parameters $\{\lambda_j^{(n)}>0 \, : \, j=1, \dots, N\}$ so that $f_j^{(n)}/\lambda_j^{(n)} \to \exp(\varphi_{\eps,j})$ in $L^1(\mathfrak m_j)$ as well as $\pi^{(n)} \to \pi_\eps$ in $L^1(\sigma)$. In particular, $\pi_\eps$ is optimal for \eqref{eq:multim_regularised_OT}, $(\varphi_{\eps,1}, \dots, \varphi_{\eps,N})$ is optimal for the dual problem \eqref{eq:multim_regularised_dual_OT}, and they are linked by the formula
\begin{align}
	\pi_\eps = 
	\Big(
		\bigotimes_{i=1}^N \exp(\varphi_{\eps,i})
	\Big) \mathcal K_\eps \, .
\end{align}
Note one can write an explicit formula for each iteration $\varphi_j^{(n)}$ in terms of the previous ones, in the same spirit of \eqref{explicit}. As we are going to see, this is in constrast with the noncommutative setting, where no explicit formulas are available (which makes the study of the convergence of the algorithm more involved).  

\section{Noncommutative optimal transport}
In this section, we are going to discuss a noncommutative analog of the multimarginal optimal transport problem \eqref{eq:multim_classic_OT}. As in \cite{Feliciangeli-Gerolin-Portinale:2022}, we only focus on the finite dimensional case, i.e. a noncommutative version of the discrete optimal transport problem as in \eqref{eq:discrete_classic_OT}.

A short discussion about the notation is due:  for a given Hilbert space $\mathfrak{g}$, we denote by $\cM(\mathfrak g)$ the set of all bounded operators over $\fg$, by $\cS(\fg)$ the hermitian elements of $\cM(\fg)$, and by $\cS_\geq(\fg)$ (resp. $\cS_>(\fg)$) the set of all the positive semidefinite (resp. positive definite) elements of $\cS(\fg)$. With a slight abuse of notation, we denote by $\Tr$ the trace operator on $\cM(\fg)$ for any Hilbert space $\fg$.  Furthemore, we denote by $\DM(\fg)$ the set of \textit{density matrices} over $\fg$, namely the elements of $\cS_\geq(\fg)$ with trace one. We also set $[N]:= \{1, ... , N\}$.

Instead of considering transport problems in metric spaces, the noncommutative framework consists of $N$ complex Hilbert spaces $\mathfrak{h}_j$  of dimension $d_j<\infty$, for $j \in [N]$, which physically represent the state spaces of $N$ subsystems. The state space of the composite system is given by $\mathfrak{h}:=\mathfrak{h}_1 \otimes \mathfrak{h}_2 \otimes \dots \otimes \mathfrak{h}_N$, which is also a finite dimensional Hilbert space of dimension $d=d_1 \cdot d_2 \cdot \dots {\cdot d_N}$, and plays the role of $\bfX$ in the classical picture. Concerning the other quantities involved in \eqref{eq:multim_classic_OT}, we have that:
\begin{enumerate}
	\item The role of the cost function $c:\bfX \to \R$ is played by a \textit{Hamiltonian} $\H \in \cS(\fh)$. 
	\item The set of probability measures over $\bfX$ are replaced by density matrices $\DM(\fh)$. In particular, having total mass equals one becomes having trace one.
	\item The notion of marginal via the projection operators $\text{Proj}_{X_i}$ is replaced by the one of \textit{reduced density matrix}. We say that $\Gamma \in \DM(\fh)$ has marginals $\bm \gamma := (\gamma_1, .. , \gamma_N)$ if 
	\begin{align}
		\label{eq:marginals_NC}
		\Tr( \gamma_i A ) = \Tr \Big( \Gamma \big(  \unit \otimes ... \otimes \unit \otimes \underbrace{A}_{\text{i-th}} \otimes \unit .. \otimes \unit  \big) \Big), \quad \forall A \in \cM(\fh_i).	
	\end{align}
	We use the short-hand notation $\Gamma \mapsto \bm \gamma$. In general, if $\Gamma$ satisfies \eqref{eq:marginals_NC} for a given $i$, we write $\tP_i(\Gamma) = \gamma_i$. Here $\tP_i: \DM(\fh) \to \DM(\fh_i)$ is the \textit{$i$-th partial trace operator}, which plays the same role of the projection  $(\text{Proj}_{X_i})_{\#}(\cdot)$ in the commutative setting.
\end{enumerate} 
The noncommutative analog of the multimarginal transport problem \eqref{eq:multim_classic_OT} is given by 
\begin{align}	\label{eq:multim_quantum_OT}
	\inf \left\lbrace \Tr(\H\Gamma) \suchthat \Gamma\in\DM(\fh) \text{ and } \Gamma\mapsto \bm \gamma  \right\rbrace \, ,
\end{align}
for some given marginal operators (or reduced density matrices) $\{ \gamma_i \in \DM(\fh_i) \}_{i \in [N]}$.

Note that without any constraint on the marginals, the optimal density matrix is a pure state (i.e. rank one projection) onto the eigenspace of $\H$ corresponding to its minimal eigenvalue. More precisely, using the spectral theorem on $\H$, we can always write 
\begin{align}
	\H = \sum_{i=1}^d \lambda_i |\psi_i \rangle \langle \psi_i|
		\, , \quad \text{where} \ 
	\sigma(\H) = \{ \lambda_i \}_{i=1}^d 
		\  \text{is the spectrum of }\H \, ,
\end{align}
where $\psi_i \in \fh$ are the eigenfunctions of $\H$ (normalised to have norm one, i.e. $\langle \psi_i | \psi_i \rangle = 1$, for every  $i$).
We are using the bracket notation $|\psi_i \rangle \langle \psi_i|$ to denote the rank-one projection onto the linear subspace spanned by $\psi_i$. The solution to the unconstrained problem of minimising $\Tr(\Gamma\H)$ is then given by the pure state $\Gamma = |\psi_{i_0} \rangle \langle \psi_{i_0}|$, where $i_0$ corresponds to the minimal eigenvalue $\lambda_{i_0} = \min_i \lambda_i$ of $\H$. This concentration phenomenon is close in spirit to what discussed in Remark~\ref{rem:euclidean_concentration}, which is typical of the unregularised setting.  In general, for given marginals $\bm \gamma$, the rank-one projection onto the span of $\psi_{i_0}$ might not be admissible.

\subsubsection*{Noncommutative entropic optimal transport}
From a physical perspective, one can study the system at zero or at positive temperature. It turns out that the quantum problem at positive temperature $\varepsilon>0$ corresponds to a regularised version of the noncommutative optimal transport problem as introduced in \eqref{eq:multim_quantum_OT}. The regularisation is performed once again by adding an entropy term, which penalises concentration, namely for which pure states are not optimal anymore. For a given reference positive operator $\bm m \in \cS_>(\fh)$, the relative entropy as introduced in \eqref{eq:def_entropy} is here replaced by  (minus) the \textit{Umegaki relative entropy} of $\Gamma$, which is given by
\begin{align}
	S(\Gamma|\bm m):= \Tr(\Gamma( \log \Gamma- \log \bm m) )
		\, , \quad 
	\Gamma \in \DM(\fh) \, .
\end{align}
In the simpler case when $\bm m = \unit$, we obtain (minus) the \textit{Von Neumann entropy} functional. Using the spectral decomposition of $\Gamma$, we can compute it as
\begin{align}
	S(\Gamma):= S(\Gamma|\unit) = \sum_{i=1}^d \lambda_i \log \lambda_i 
		\, , \quad 
	\text{if} \quad 
		\Gamma = \sum_{i=1}^d \lambda_i |\psi_i \rangle \langle \psi_i| 
	\, ,
\end{align}
which shows that $S(\Gamma)$ can be interpreted as the entropy functional \eqref{eq:def_entropy} evaluated in the empirical measure with support on the spectrum of $\Gamma$ relative to the counting measure.

In general, we assume $\bm m$ to be of product form, that is $\bm m = \otimes_{i=1}^N m_i$, for some $\{ m_i \in \cS_>(\fh_i) \}$. The entropic noncommutative optimal transport problem with parameter $\eps >0$ is then given by
\begin{align}
		\label{eq:multim_quantum_regularised_OT}
	\primin_{\bm m}(\bm \gamma):=
	\inf \left\{
		\Tr(\H\Gamma)+\varepsilon S(\Gamma|\bm m)
		\suchthat \Gamma\in\DM(\fh) \text{ and } \Gamma\mapsto \bm \gamma
	\right\} \, ,
\end{align} 
for a given set of marginals $\{ \gamma_i \in \DM(\fh_i) \}_{i \in [N]}$.

\begin{rem}[Many-body problems]
	Typically in quantum many-body problems, the Hamiltonian $\H$ to which the whole system is subject is of the form $\H=\H_0+\H_{\text{int}}$, where $\H_0$ is the non-interacting part of the Hamiltonian, i.e.  $\H_0=\bigoplus_{j=1}^N \H_j:=\H_1 \otimes \unit \dots \otimes \unit+ \unit\otimes \H_2 \otimes \unit \dots \otimes \unit +\dots +\unit\otimes \dots \otimes \unit \otimes \H_N$, with $\H_j$ acting on $\mathfrak{h}_j$, and $\H_{\text{int}}$ is its interacting part. Now, suppose we have knowledge of the states $\bm \gamma=(\gamma_1, \dots, \gamma_N)$ of the $N$ subsystems.	
	Then the energy of the composite system at temperature $\varepsilon > 0$ is given by
	\begin{align}
		\label{eq:multimarginalGSEposT}
		\inf_{\Gamma \mapsto \bm \gamma} \left\{\Tr(\H\Gamma)+\varepsilon S(\Gamma)\right\}=\sum_{j=1}^N\Tr(\H_j \gamma_j)+\inf_{\Gamma \mapsto \bm \gamma} \left\{\Tr(\H_{\text{int}}\Gamma)+\varepsilon S(\Gamma)\right\} \, ,
	\end{align}	
where we clearly see in the second term of the right-hand side an entropic problem  \eqref{eq:multim_quantum_regularised_OT}.
\end{rem}

\begin{rem}[Lack of concentration]
	Denote by $\bar \Gamma_\eps \in \DM(\fh)$ the density matrix which minimises the energy in \eqref{eq:multim_quantum_regularised_OT} with $\bm m = \unit$  \textit{without the marginals constraints}. We claim that $\bar \Gamma_\eps$ is never a pure state; in fact, we shall show that $\bar \Gamma_\eps$ has trivial kernel. Indeed, suppose that $\ker \bar \Gamma_\eps \neq \{0\}$ and denote by $\Pi_{\ker \bar \Gamma_\eps}$ the projection onto it. Define $\tilde \Gamma_s := \bar \Gamma_\eps + s \Pi_{\ker \bar \Gamma_\eps}$ for $s >0$, and observe that $\tilde \Gamma_s$ has the same eigenspaces of $\bar \Gamma_\eps$ but eigenvalue $s>0$ for every element in $\ker \bar \Gamma_\eps$. Computing the entropy, we see that
	\begin{align*}
		S(\tilde \Gamma_s) = S(\bar\Gamma_\eps) + s \log s \, d_{\ker} 
		\, , \quad 
		\text{where} \quad d_{\ker} := \dim{\ker \bar \Gamma_\eps} \, .
	\end{align*}	
	Therefore, by construction we obtain that
	\begin{align*}
		\Tr (\H \tilde \Gamma_s) -\eps S(\tilde \Gamma_s)
		&= 
		\Tr (\H \bar \Gamma_\eps) - \eps S(\bar \Gamma_\eps)
		+ s 
		\left(
		\Tr (\H \Pi_{\ker \bar \Gamma_\eps}) 
		- \eps \log s \, d_{\ker}
		\right) 
\\
	&>
		\Tr(\H \bar \Gamma_\eps) - \eps S(\bar \Gamma_\eps) 
	\end{align*}	
	as soon as $s < \exp \left( (\eps d_{\ker})^{-1} \Tr( \H \Pi_{\ker \bar \Gamma_\eps} )\right) \in \R_+$, which contradicts the optimality of $\bar \Gamma_\eps$.
\end{rem}

\subsubsection*{Dual formulation in the noncommutative setting}
Another common feature between the classical and quantum setting is the existence of an equivalent, dual formulation of the problem \eqref{eq:multim_quantum_regularised_OT}. To simplify the discussion, we assume from now on that $\ker \gamma_i = \{ 0 \}$, for every $i \in [N]$. In the general case, it suffices to consider the restriction to the set $	\mathcal O:= \bigotimes_{i=1}^N \big( \ker \gamma_i \big)^\perp $, see \cite[Remark~3.9]{Feliciangeli-Gerolin-Portinale:2022}.
In this case, the corresponding dual problem is defined as
\begin{gather}	\label{eq:multim_quantum_regularised_dual_OT}
	\dualin_{\bm m}(\bm \gamma) = \sup \Bigg\{  \sum_{i=1}^N \Tr(U_i \gamma_i)-\eps \Tr \bigg( \exp\bigg[ \frac{\bigoplus_{i=1}^N U_i - \H_{\bm m}}{\eps} \bigg] \bigg) \suchthat U_i\in \cS(\fh_i) \Bigg\} + \eps  \, , \quad 
\end{gather}
where $\H_{\bm m} := \H - \eps \log \bm m$ and $\bigoplus$ denotes the Kronecker sum of operators, defined as
\begin{align}
	\bigoplus_{i=1}^N U_i := 
	\sum_{i=1}^N
		\big(  \unit \otimes ... \otimes \unit \otimes \underbrace{U_i}_{\text{i-th}} \otimes \unit .. \otimes \unit  \big) \, .
\end{align}

The first result proved in \cite{Feliciangeli-Gerolin-Portinale:2022} is the following duality result.
\begin{theo}[Duality]
	For any given $\bm \gamma =( \gamma_i \in \DM(\fh_i) )_{i \in [N]} $, $\H \in \cS(\fh)$, we have that
	\begin{itemize}
		\item[(i)] the primal and dual problems coincide, i.e. $\primin_{\bm m} (\bm \gamma) = \dualin_{\bm m}(\bm \gamma)$.
		\item[(ii)] there exists a maximiser $(   U_i^\ep \in \cS^{\hat d_i} )_{i=1}^N$ in \eqref{eq:multim_quantum_regularised_dual_OT}, which is unique up to trival translation. More precisely, if $( \tilde U_i^\ep \in \cS^{\hat d_i} )_{i=1}^N$ is another maximiser, then $\tilde U_i^\ep - U_i^\ep = \alpha_i \unit$ with $\alpha_i \in \R$ and $\sum_i \alpha_i =0$.
		\item[(iii)] There exists a unique $ {\Gamma}^{\ep}\in \DM^{\bm d}$ with $ {\Gamma}^{\ep} \mapsto \bm \gamma$ minimiser in \eqref{eq:multim_quantum_regularised_OT}. Moreover, $ {\Gamma}^{\ep}$ and $(   U_i^\ep )_i$ are related via the formula
		\begin{align}	\label{eq:optimality_maintheorem}
			{\Gamma}^{\ep}=\exp\left( \frac {\bigoplus_{i=1}^N   U_i^\ep-\H_{\bm m}}{\ep}\right)  \, .
		\end{align}		
	\end{itemize}  
\end{theo}
One should compare the shape of the optimisers given in \eqref{eq:optimality_maintheorem} with the explicit formula given in \eqref{eq:optimiser_formula}. The main difference in the noncommutative setting is that, a priori, we cannot write the optimiser in product form. Indeed, we have that
\begin{align}
	\exp(A+B) = \exp(A) \exp(B) 
		\qquad
	\text{if}
		\qquad 
	[A,B] := AB - BA = 0 \, ,
\end{align}
but it is not clear whether the optimal potentials $U_i^\eps$ commute (i.e. $[U_i^\eps, \H]=0$) with the Hamiltonian $\H$. This observation is crucial when one wants to construct a noncommutative version of the Sinkhorn algorithm: we do not expect anymore to be able to write explicit iterations as in \eqref{explicit}, hence also direct manipulations are not available anymore, already at the level of the definition of a noncommutative notion of the $(c,\eps)$-transform.

\subsubsection*{A noncommutative $(\H,\eps)$-transform and the Sinkhorn algorithm}
In this section, we discuss the noncommutative version of the $(c,\eps)$-transform as introduced in \cite{Feliciangeli-Gerolin-Portinale:2022}. For simplicity, we discuss the case $\bm m = \unit$, the general problem following by simply considering the modified Hamiltonian $\H_{\bm m}^\eps:= \H - \eps \log \bm m$, see also \cite[Remark~2.5]{Feliciangeli-Gerolin-Portinale:2022}.
We denote by $\dualf$ the maximising functional of the dual problem, namely for every $\bm U=(U_1,\dots,U_N)\in  \bigtimes_{i=1}^N \cS(\fh_i)$, we set
\begin{align}
		\dualf(\bm U):=
	\sum_{i=1}^N \Tr(U_i \gamma_i)-\ep \Tr \bigg( \exp\bigg[ \frac{\bigoplus_{i=1}^N U_i - \H}{\ep} \bigg] \bigg) + \eps \, .
\end{align}
The $j$-th \textit{$(\H,\eps)$-transform} of $\bm U$ is then the solution to the variational problem 
\begin{align}	\label{eq:max_Heps_transf}
	\Argmax \Big\{ \dualf(U_1, \dots, U_{j-1}, V, U_{j+1}, \dots, U_N) \suchthat V \in \cS(\fh_j) \Big\}  \, .
\end{align}
One can show that the above maximisation problem admits a unique solution, that we denote by $\fT_j^\ep\big(\bm U\big) \in \cS(\fh_j)$. A direct analysis of the optimality conditions associated with the maximisation \eqref{eq:max_Heps_transf} shows that $\fT_j^\ep\big(\bm U\big)$ can be characterised as the unique solution to 
\begin{align}
	\tP_j \left[ \Sink(\bm U)  \right] = \gamma_j
		\quad 
	\text{for} 
		\quad 
	\Sink(\bm U):= \exp\left(\frac {\bigoplus_{i=1}^{j-1} U_i \oplus \fT_j^\eps(\bm U) \bigoplus_{i=j+1}^N U_i-\H}{\ep}\right) \, ,
\end{align}
where recall that $\tP_j$ denotes the $j$-th marginal operator \eqref{eq:marginals_NC}.
In other words, the density matrix $\Sink(\bm U) \in \DM(\fh)$ has the right $j$-th marginal operator $\gamma_j$, similarly to the commutative setting. On the other hand, a crucial difference  is that, due to the possible lack of commutativity between the involved operators, one has no way to provide an explicit formula for $\fT_i^\ep\big(\bm U\big)$. Nonetheless, a careful analysis of the maximisation problem still provides uniform bounds in terms of spectral bounds for the Hamiltonian $\H$. More precisely, in \cite{Feliciangeli-Gerolin-Portinale:2022} the authors show that there exists a \textit{renormalisation operator} $\ren: \bigtimes_{i=1}^N \cS(\fh_i) \to \bigtimes_{i=1}^N \cS(\fh_i)$ with the following properties:
\begin{enumerate}
	\item $\ren$ is an \textit{idempotent operator}, i.e. $\ren \circ \ren = \ren$.
	\item $\ren(\bm U) = \bm U + \big(\alpha_1(\bm U) \unit, \dots, \alpha_N (\bm U) \unit\big)$, where $\alpha_i(\bm U) \in \R$ is so that
	\begin{align}	\label{eq:alphas}
		\sum_{i=1}^N \alpha_i(\bm U) = 0
			\, , \quad
		\text{for every}
			\quad
		\bm U \in \bigtimes_{i=1}^N \cS(\fh_i).
	\end{align}
	Therefore, we have that $\bigoplus_i (\ren(\bm U))_i = \bigoplus_i U_i$. In particular, it does not change the value of the dual functional, namely $\dualf(\ren (\bm U) ) = \dualf(\bm U)$.
	\item Once we compose $\ren$ with the $(\H,\eps)$-transform operators, we obtain a \textit{bounded} nonlinear map from $\bigtimes_{i=1}^N \cS(\fh_i)$ onto itself. More precisely, we have that
	\begin{align}	\label{eq:uniform_bound}
		\big| (\ren \tau(\bm U))_i - \eps \log \gamma_i   \big|	\leq 2 \| {\rm H} \|_{\infty}  \unit, \quad \forall i \in [N] \, ,
	\end{align}
	where $\tau$ denotes the \textit{Sinkhorn operator}, given by
	\begin{align}	\label{eq:Sinkhorn_operator}
		\begin{gathered}
			\tau: \bigtimes_{i=1}^N \cS(\fh_i) \to \bigtimes_{i=1}^N \cS(\fh_i) , \\
			\tau(\bm U) := (\Sink_N \circ \dots \circ \Sink_1) (\bm U) \, .
		\end{gathered}
	\end{align}
\end{enumerate}

There are two important observations worth discussion at this point: the first is that a renormalisation procedure is necessary in order to obtain uniform bounds of the type \eqref{eq:uniform_bound}. Indeed, the dual functional being invariant by translation of operators as in \eqref{eq:alphas} implies that no uniform bound is in general expected on maximising sequences for $\dualf$. Secondly, an interesting feature of the bound in \eqref{eq:uniform_bound} is that it is \textit{dimension free}, in the sense that the right-hand side only depends on $\H$ but not on $\bm d$. This possibly represents a hint to the fact that some of the estimates and properties of the $(\H,\eps)$-transform obtained in \cite{Feliciangeli-Gerolin-Portinale:2022} in the finite dimensional setting ($\bm d < \infty$) might be extendable to an infinite dimensional setting. Of course, in this case the estimate would be useful only for bounded operators $\H$. In case of unbounded operators, a different type of analysis would be certainly required.

We are finally ready to introduce the noncommutative version of the Sinkhorn algorithm. Its main goal is to provide a sequence of density matrices $\{ \Gamma^{(n)} \}_n$ having one correct marginal at a time, in such a way that $\Gamma^{(n)} \to \Gamma^\eps$ (the optimal solution) as $n \to \infty$.
In contrast with the commutative framework, no explicit formulas are available, hence no product form for $\Gamma^{(n)}$is ensured. On the other hand, we use the $(\H,\eps)$-transforms in order to ensure the correct marginals (one at a time) and the renormalisation operator $\ren$ to obtain the sought compactness to pass to the limit in the number of iterations going to $\infty$. 

\noindent
{\underline{Initialisation}}. \ We initialise 
$
\Gamma^{(0)}= \exp 
	\left(
		\frac{ \bigoplus_{i=1}^N U_i^{(0)} - \H}{\eps}
	\right) 
\in \DM(\fh)
$ in exponential form, for a given, initial vector of self-adjoint operators $\bm U^{(0)}= (U_1^{(0)}, \dots, U_N^{(0)})$, with $U_i^{(0)} \in \cS(\fh_i)$.

\noindent 
{\underline{Iteration}}. \ We iteratively define
$
\Gamma^{(n)}= \exp 
	\left(
		\frac{ \bigoplus_{i=1}^N U_i^{(n)} - H}{\eps}
	\right) 
\in \DM(\fh)
$
by applying the $(\H,\eps)$-transforms for each component $i \in [N]$. Precisely, we iteratively define
\begin{align*}
	\bm U^{(n)} := \tau \big( \bm U^{(n-1)} \big)
		\, , \quad 
	n \in \N \, ,
\end{align*}
where $\tau$ is the Sinkhorn operator introduced in \eqref{eq:Sinkhorn_operator}. Taking advantage of the uniform estimates for the renormalisation operator \eqref{eq:uniform_bound}, one gets the following convergence result.

\begin{theo}\emph{(\cite[Prop.~5.1]{Feliciangeli-Gerolin-Portinale:2022})}
The previous described algorithm converges as $n \to \infty$, in the sense that: 	
\begin{enumerate}
		\item There exist $\bm U^\eps \in \bigtimes_{i=1}^N \cS(\fh_i) $ and $\bm \alpha^n \in \bR^{N}$ with $\sum_{i=1}^N \alpha_i^n =0$ such that
		\begin{align}
			\bm U^{(n)}+ \bm \alpha^n \rightarrow \bm U^\ep \quad \text{as } n \to \infty \, .
		\end{align}
		\item $\bm U^\ep = (\bm U_1^\ep, \dots, \bm U_N^\ep)$ is optimal for the dual problem $\dual(\gamma)$ as defined in \eqref{eq:multim_quantum_regularised_dual_OT}.
		\item $\Gamma^{(n)}$ converges as $n \to \infty$ to some $\Gamma^\eps \in \DM(\fh)$ which is optimal for the primal problem $\prim(\bm \gamma)$ as defined in \eqref{eq:multim_quantum_regularised_OT}. In particular, it holds
		\begin{align}	\label{eq:optim_cond_sect5}
			\Gamma^\eps=\exp\left(\frac {\bigoplus_{i=1}^N U^\eps_i-\H}{\eps}\right) \, .
		\end{align}	
	\end{enumerate}
\end{theo}
\noindent
Note that the renormalisation with $\bm \alpha^n$ leaves $\Gamma^{(n)}$ invariant, i.e. 
\begin{align*}
	\Gamma^{(n)}= \exp 
	\left(
		\frac{ \bigoplus_{i=1}^N U^{(n)}_i-\H}{\eps}
	\right) 
	= 
	\exp
	\left(
		\frac{ \bigoplus_{i=1}^N \big( U^{(n)}_i + \alpha_i^n \big) -\H}{\eps}
	\right) 
	\, ,
\end{align*}
therefore the convergence of $\bm U^{(n)} + \bm \alpha^n$ ensures the convergence of $\Gamma^{(n)}$. 
\subsubsection*{The symmetry-constrained problems: fermionic and bosonic setting}
It is interesting to see that the theory presented up to this point can be also extended in the physically relevant case where additional symmetry constraints on the space of density matrices are added. 
Two typical types of particles in physics are \textit{bosons} and \textit{fermions}, which from a mathematical perspective are described by density operators that are respectively symmetric and antisymmetric with respect to the underlined product space. Particles are hereby considered indistinguishable, in particular the Hibert spaces $\{\fh_i\}_i$ are to be considered the same. In the finite dimensional setting, this can be identified with the complex space $\bC^d$ for a given dimension $d \in \N$. 

From a mathematical perspective, we shall consider the symmetric and antisymmetric tensor product of the one-particle Hilbert space (instead of the free tensor product in the setting without symmetries), namely we work with
\begin{align}
	\left(\bigodot_{i=1}^N \bC^d\right)
		\, , \,
	\left(\bigwedge_{i=1}^N \bC^d\right)
		\subset \bigotimes_{i=1}^N \bC^d 
		\, ,
\end{align}
where $\odot$ (resp. $\wedge$) denotes the symmetric (resp. antisymmetric) tensor product.
Note that the cardinality of $\bigwedge_{i=1}^N \bC^d$ is $\binom{d}{N}$ if $d \geq N$, and $\bigwedge_{i=1}^N \bC^d=\{0\}$ if $d < N$, due to the fact that $e \wedge e = 0$.  In particular, in the fermionic setting, we assume to work with a number of particles $N$ which is smaller than the considered dimension $d$.
We denote by 
\begin{align} \label{eq:bosonicfermionicDM}
	\DM_+^{\bm d}:=\DM\left(\bigodot_{i=1}^N \bC^d\right), \, \   \DM_-^{\bm d}:=\DM\left(\bigwedge_{i=1}^N \bC^d\right) \, \ \ \subset \, \DM^{\bm d} 
		:= \DM
		\left(
			\bigotimes_{i=1}^N \bC^d 
		\right) \, ,
\end{align}
the set of bosonic and fermionic density matrices. To simplify the notation, we also write $\cM^d := \cM(\bC^d)$, $\cS^d := \cS(\bC^d)$. It is key to observe that, in both cases, the marginals operators of a given density matrix $\Gamma \in \DM_{\pm}^d$ all coincide.
 
The fermionic symmetry constraint has an important consequence on the type of operators that one can obtain as marginals, namely not every operator $\gamma \in \DM^d$ can be obtained as marginal of a fermionic operator. It is indeed well-established that the existence of a $\Gamma \in \DM_-^{\bm d}$ so that $\Gamma \mapsto \gamma$ (where $\Gamma \mapsto \gamma$ means that $\Gamma$ has all marginals equal to $\gamma$) is equivalent to $\gamma$ satisfying the so-called \textit{Pauli exclusion principle}, i.e. if and only if $\gamma \in \DM^d$ and $\gamma\leq 1/N$ (see for example \cite[Theorem 3.2]{LiebSeiringer10}). As we are going to see in the final part of these notes, the dual formulation of the entropic transport problem provides, in a natural way, a more variational interpretation of the Pauli principle.

In order to talk about optimal transport, we must fix a reference Hamiltonian. In the symmetric setting, it is natural to consider an operator which also satisfies some additional symmetry conditions. To this purpose, for $i \in [N]$, we introduce the \textit{permutation operators} $\tS_i : \cM^{\bm d} \simeq \bigotimes_{j=1}^N \cM^{d_j} \to \cM^{\bm d}$ as the map defined by 
\begin{align*}
	\tS_i \bigg( \bigotimes_{j=1}^N A_j \bigg) = A_1 \otimes \dots \otimes A_{i-1} \otimes A_{i+1} \otimes \dots \otimes A_N \otimes A_i,
\end{align*}
for any $A_j \in \cM^{d_j}$ and extended to the whole $\cM^{\bm d}$ by linearity. The natural assumption on $\H \in \cS^{\bm d}$ is then to assume that 
\begin{align}	\label{eq:symmetry}
	\tS_i \circ \H \circ \tS_i = \H \, , \quad \forall i = 1, \dots, N \, .
\end{align}
An interpretation of the latter condition is that the quantum systems is driven by an Hamiltonian which does not distinguish between the different $N$ particles.

We are now in position to define the bosonic and the fermionic entropy-regularised optimal transport problem. For any $\gamma \in \DM^d$, we define the \textit{bosonic primal problem} as 
	\begin{align}	\label{eq:primal_bosons}
		\prim_+(\gamma) := \inf \left\lbrace \Tr(\H\Gamma)+ \ep  \Tr(\Gamma \log \Gamma) \suchthat \Gamma\in\DM_+^{\bm d} \text{ and } \Gamma\mapsto \gamma  \right\rbrace \, .
	\end{align}
In light of the Pauli principle for fermionic density operators, we define the \textit{fermionic primal problem} for a given $\gamma \in \DM^d$ satisfying $\gamma \leq 1/N$ as 
	\begin{align}	\label{eq:primal_fermions}
		\prim_-(\gamma) := \inf \left\lbrace \Tr(\H\Gamma)+ \ep  \Tr(\Gamma \log \Gamma) \suchthat \Gamma\in\DM_-^{\bm d} \text{ and } \Gamma\mapsto \gamma  \right\rbrace \,.
	\end{align}
In order to introduce and discuss the dual formulation of the bosonic/fermionic primal problems, we first introduce two natural projection operators $\Pi_+$ (resp. $\Pi_-$), given by 
\begin{align}\label{eq:projectionFermionicBosonic}
	\Pi_+: \bigotimes_{i=1}^N \bC^d \to \bigodot_{i=1}^N \bC^d  \, ,  \qquad 	\Pi_-:\bigotimes_{i=1}^N \bC^d \to \bigwedge_{i=1}^N \bC^d   \, .
\end{align} 
For any given operator $A \in \cS^{\bm d}$, we denote by $A_\pm$ the corresponding projection onto the symmetric space, obtained as $A_\pm := \Pi_\pm \circ A \circ \Pi_\pm$ .
Consequently, for any $\gamma \in \DM^d$, we define the \textit{bosonic  dual functional} $\fbD$ and the \textit{fermionic dual functional} $\ffD$ as 
	\begin{align}	\label{eq:primal_fermions_bosons}
		\fbfD: \cS^d \to \R \, , \quad \fbfD(U) := \Tr( U \gamma ) - \ep
		\Tr \left( 
		\exp \bigg[ \frac1\ep \bigg(\frac1N  \bigoplus_{i=1}^N U - H \bigg)_\pm  \bigg]
		\right) + \ep \, .
	\end{align}
	The corresponding dual problems are given by
	\begin{align}	\label{eq:dual_fermions_bosons}
		\bfD(\gamma) := \sup \left\{ \fbfD(U) \suchthat U \in \cS^d \right\} \, .
	\end{align}
Note that a priori $\fD(\gamma)$ is well-defined for $\gamma \in \DM^d$ without any additional assumption, whereas $\prim_-(\gamma)$ is only well defined for $\gamma\in \DM^d$ such that $\gamma\leq 1/N$, due to the presence of the Pauli principle. Nonetheless, once we try to maximise the dual functional, the constraint provided by the Pauli principle naturally arises as an admissibility condition in order to have $\fD(\gamma) < \infty$. Another interesting feature concerns the existence of maximisers: to ensure the existence of a maximiser for $\ffD$, one has to further impose that $\gamma<1/N$, a condition that the authors in \cite{Feliciangeli-Gerolin-Portinale:2022} call \textit{strict Pauli principle}. This variational interpretation of the principle based on the dual formulation is the content of the next theorem  \cite[Prop.~2.8]{Feliciangeli-Gerolin-Portinale:2022}.
\begin{proposition}[Pauli's principle and duality]		\label{prop:pauli}
	Let $N \geq d$. The following are equivalent:
	\begin{enumerate}
		\item $
		\fD(\gamma) < + \infty
		$ if and only if $\gamma \in \DM^d$ and $\gamma \leq \frac1N$.
		\item There exists a maximizer $U_0 \in \cS^d$ of $\emph{D}_\gamma^{-,\eps}$ if and only if $\gamma \in \DM^d$ and  $0 < \gamma < \frac1N$. 
	\end{enumerate}
\end{proposition}
The proof of the necessity of the Pauli principle to have the dual functional bounded from above provided in \cite{Feliciangeli-Gerolin-Portinale:2022} is constructive. In particular, reasoning by contradiction, one can explicitly construct a sequence of operators $U^n \in \cS^d$ along which the functional $\text{D}_\gamma^{-,\eps}$ diverges to $+\infty$. 
To see this, suppose by contradiction that the Pauli principle is not satisfied. By spectral calculus, and up to reordering the terms in the sum, we can without loss of generality assume that 
	\begin{align*}
		\gamma = \sum_{i=1}^d \gamma_i |\psi_i \rangle \langle \psi_i|, \quad \gamma_1 > \frac1N \, .
	\end{align*}
The sought sequence of operators $U^n \in \cS^d$ is then obtained as matrices having the same eigenspaces of $\gamma$, choosing a \textit{very positive} eigenvalue in correspondence to first eigenspace (where the condition described by the Pauli principle does not hold), and negative eigenvalues otherwise. Precisely, we define
	\begin{align}	\label{eq:defUn}
		U^n := \sum_{i=1}^d u_i^n |\psi_i \rangle \langle \psi_i|, \quad u_1^n := n \, , \quad u_j^n := -\frac{n}{N-1} , \quad  \forall j \geq 2 \, .
	\end{align}
	This sequence is constructed in such a way that, for any choice of indexes $\bm j = (j_1, \dots, j_N)$ with $j_i \in \{1, \dots , d\}$ \textit{without repetition} (namely $j_i \neq j_k$ if $i \neq k$),  we have that
	\begin{align*}
		 \exp \left( \frac1N \sum_{i=1}^N u_{j_i} \right) 
		\begin{cases}
			= 1 & \text{if } j_i = 1 \text{ for some }i \,  ,\\
			\leq 1 & \text{otherwise} \, .
		\end{cases}
	\end{align*}
	This ensures that the nonlinear part of $\ffd(U)$ can be estimated from above  by the dimension of the antisymmetric product space. In particular, we have that
	\begin{align}	\label{eq:lb}
		\ffd(U^n) \geq  \sum_{j=1}^d \gamma_j u_j^n - C \binom{d}{N} \, .
	\end{align}
	The claim is that the linear part in the bound above goes to $+\infty$ as $n \to +\infty$. Indeed
	\begin{align}	\label{eq:linear_example}
		\sum_{j=1}^d \gamma_j u_j^n = n \bigg(  \gamma_1 - \frac{1}{N-1} \sum_{i=2}^d \gamma_i \bigg) = \frac{n}{N-1} \Big(N \gamma_1 - 1 \Big) \, , 
	\end{align}
	where we used that $\sum_i \gamma_i = 1$. From this, using $\gamma_1 > \frac1N$ and \eqref{eq:lb}, we deduce $\ffd(U^n) \to +\infty$ as $n \to +\infty$, which provides the sought contradiction. 

Finally, putting together the duality result in the non-symmetric case and the relation to the Pauli principle, one can prove a duality result also in the bosonic and fermionic setting, see \cite[Thm~2.9]{Feliciangeli-Gerolin-Portinale:2022}.
\begin{theo}[Fermionic and bosonic duality] \label{theo:dualBosFerm}
	Let $\H \in \cS^{\bm d}$ satisfy the condition \eqref{eq:symmetry}. 
	\begin{itemize}
		\item[(i)]  Assume that $N \geq d$. For any $\gamma \in \DM^d$ such that $\gamma \leq \frac1N$, the fermionic primal and dual problems coincide, i.e. $\prim_-(\gamma) = \fD(\gamma)$. If additionally $\gamma$ satisfies the strict Pauli principle $0 < \gamma < \frac1N$, then $\emph{D}_\gamma^{-,\eps}$ admits a unique maximiser $U_-^\ep \in \cS^d$ and
		\begin{align}	\label{eq:optimum_primal_fermions}
			\Gamma_-^\ep = \exp \left( \frac1\ep \bigg[ \frac1N \bigoplus_{i=1}^N U_-^\ep - \H  \bigg]_- \right) 
		\end{align}
		is the unique optimal fermionic solution to the primal problem $\prim_-(\gamma)$.
		\item[(ii)] Fix $ N \in \N$. For any $\gamma \in \DM^d$, the bosonic primal and dual problems coincide, i.e. $\prim_+(\gamma) = \bD(\gamma)$. Moreover, if $\gamma >0$ (or equivalently $\ker \gamma = \{ 0 \}$), then $\emph{D}_\gamma^{+,\eps}$ admits a unique maximiser $U_+^\ep$ and
		\begin{align}	\label{eq:optimum_primal_bosons}
			\Gamma_+^\ep = \exp \left( \frac1\ep \bigg[ \frac1N \bigoplus_{i=1}^N U_+^\ep - \H  \bigg]_+ \right) 
		\end{align}
		is the unique minimiser in the primal problem $\prim_+(\gamma)$.
	\end{itemize}
\end{theo}

\subsection*{Acknowledgments}
We are very thankful to the organisers of the School and Workshop
on Optimal Transport on Quantum Structures at Erd\"os Center: J. Maas,
S. Rademacher, T. Titkos, D. Virosztek, for the nice scientific experience. We are grateful to the anonymous reviewer for their careful reading and useful suggestions. The author also acknowledges fundings from the Deutsche Forschungsgemeinschaft (DFG, German Research Foundation) under Germany's Excellence Strategy
- GZ 2047/1, Projekt-ID 390685813.

%
\end{document}